\documentstyle[preprint,eqsecnum,aps]{revtex}
\def\be{\begin{equation}} 
\def\ee{\end{equation}} 

\def\centereps#1#2#3{\vskip#2\relax\centerline{\hbox to#1{\special
  {eps:#3 x=#1, y=#2}\hfil}}}

\preprint{HEP/123-qed}

\def\pa{\subparagraph} 
\begin{document}   

\title{ Scaling for Mixtures of Hard Ions and Dipoles in the Mean Spherical Approximation}

\author{L. Blum }

\address{Department of Physics P.O. Box 23343, University of
Puerto Rico, Rio Piedras, PR 00931-3343 }

\date{\today}
\maketitle
\begin{abstract}

Using  new  scaling parameters $\beta_i$, we derive  simple  expressions for the excess thermodynamic properties of the  Mean Spherical Approximation (MSA) for the ion-dipole mixture. For the MSA and its extensions we have shown that the thermodynamic excess functions are  a function of a reduced set of scaling matrices ${\mathbf\Gamma}_\chi$. We show now that for factorizable interactions like the hard ion-dipole mixture there is a further reduction to a diagonal matrices ${\mathbf\beta}_\chi$. The excess thermodynamic properties are simple functions of these new parameters. For the  entropy we get 
\[
S=-\left\{ \frac{k V}{3 \pi}\right\}\left( {\cal F}[{\mathbf\beta}_\alpha]\right)_{\alpha\in {\mathbf \chi}}
\]
where ${\cal F}$ is an algebraic functional of the  scaling matrices of irreducible representations $\chi$ of the closure of the Ornstein-Zernike.  

The new  scaling parameters $\beta_i$,  are  also simply related to  the chemical  potentials of the components. The analysis also provides a new definition of the Born solvation energy for arbitrary concentrations of electrolytes.

\end{abstract}

\pacs PACS 61.20.Gy.
\section{Introduction}\label{intro}

The remarkable simplicity of the  mean spherical approximation (MSA)\cite{blum1el,blum2,bluka1,kaho} and its extensions using Yukawa closures \cite{ubria1,ubria2,hernan1} can be summarized by the fact that for a wide class of systems the entropy  has the same simple  functional form. The MSA \cite{percus} is the solution of the  linearized Poisson-Boltzmann equation, just as the Debye-H\"uckel (DH) theory\cite{dh}. It shares with the DH theory the remarkable simplicity of a one parameter description \cite{bluma}( the screening  length $\kappa$) of all the thermodynamic and structural properties of rather diverse systems. The major difference is that in the MSA the excluded volume of all the ions are treated exactly.\\

 The MSA thermodynamic
properties of  real electrolyte solutions are expressed by
simple analytical formulas which are in many cases remarkably accurate. This is so because the MSA satisfies a number of  exact asymptotic relations, such as the  large charge, large density limits of Onsager
\cite{onsager,yr4,yr5,yr6} and  the large charge, small density limits implied in the Wertheim-Ornstein-Zernike equation. Very recently \cite {ubria1,ubria2}we have been able to extend the MSA closure analytical solutions to any arbitrary closure that can be expanded in damped exponentials (Yukawa functions), and obtain explicit, analytical forms of the excess thermodynamic functions in terms of a matrix of scaling parameters ( the EMSSAP, or Equivalent Mean Spherical Scaling Approach\cite{hernan1}).
\\

For systems with Coulomb and screened Coulomb interactions in a variety
 of mean spherical approximations (MSA) it is known that
the solution of the Ornstein Zernike (OZ) equation is given in terms of a single
screening parameter $\Gamma$. This includes the 'primitive' model of electrolytes,
in which the solvent is a continuum dielectric, but also models in which 
the solvent
is a dipolar hard sphere, and much more recently the YUKAGUA model of water \cite{yukag} that 
has  the correct tetrahedral structure.
The MSA can be deduced from a variational principle in which the energy is 
obtained from simple electrostatic considerations and the entropy is a universal
function. For the primitive model it is
$$ \Delta S= -k V\frac{\Gamma^3}{3 \pi}$$
 where $\Gamma $ is MSA the screening parameter. In general it will be of the form 
$ \Delta S= S({\mathbf \Gamma})$ 
which is independent of the geometry of the problem.
${\mathbf \Gamma}$ is a general, non diagonal, scaling matrix.
We have shown that in all known cases the
scaling matrix ${\underline{\underline \Gamma}}$ is obtained from   the variational
principle

\be
 \frac{\partial A}{\partial {\mathbf \Gamma}}=0
\label{A}
\ee

Ionic solutions are mixtures of charged particles,
the ions, and the neutral solvent particles,
most commonly water, which has an asymmetric charge
distribution, a large electric dipole and higher electric moments. Because of the 
special nature of these forces the charge distribution around a given ion and 
the thermodynamics does satisfy a series of conditions or sum rules. One 
remarkable property of mixtures of classical charged particles is that 
because of the very long range of the electrostatic forces, they must 
create a neutralizing atmosphere of counterions, which shields 
{\em perfectly} any charge or fixed charge distribution.
Otherwise the partition function, and therefore all the thermodynamic functions, will be 
divergent \cite{blmg}.  The size of the region where this charge shielding
occurs depends not only on the electrostatics, but also on all the other
 interactions of the system. For spherical ions this means:
\begin{enumerate}
\item The internal energy $E$ of the ions is always the sum of the energies 
of capacitors. For spherical ions the capacitor is a spherical capacitor, and 
the {\it exact} form of the energy is
\be
 \Delta E=-\frac{e^2}{\varepsilon} 
\sum_i\rho_{i}z_i  \frac{ z_i^*}{1/\Gamma_i+\sigma_i },
\label{3.2}
\ee 
where $z_i^*$ is the effective charge,
 $\beta = 1/kT$ is the usual Boltzmann thermal factor, $\varepsilon$ 
 is the dielectric constant, $e$ is the elementary charge, and ions $i$ 
have charge, diameter and density $ez_i$, $\sigma_i $, $\rho_i$, respectively.
For the continuum dielectric primitive model $\Gamma_i=\Gamma$ for all i.

\item The Onsagerian limits. When the ionic concentration goes to infinity and
at the same time the charge diverges, then the limiting energy is bounded by 
\be
 \Delta E=-\frac{e^2}{\varepsilon} 
\sum_i\rho_{i}z_i  \frac{ z_i^*}{\sigma_i },
\label{3.2a}
\ee 
obtained by setting $\Gamma_i\rightarrow \infty$

\item A further exact limit is the DH limiting law, which simply requires that for all ions
in the system
\be
2 \Gamma_i \rightarrow \kappa \qquad with \qquad \kappa^2=\frac{4 \pi \beta e^2}
{\varepsilon}\sum_{j=1}^m \rho_j  z_j^2.
\label{kappa}
\ee
\item Finally in systems that are strongly associating in the limit of {\it total}
association the above equation still holds. This means that if component 1 forms a n-mer the
DH limiting law must satisfy
\be
 \kappa^2=\frac{4 \pi \beta e^2}
{\varepsilon}\left[\sum_{j=2}^m \rho_j  z_j^2+ \rho_1 (n z_1)^2\right].
\label{kke}
\ee
This limiting law is  satisfied  to lowest order in the multipole expansion by the  closures of the Werteim-Ornstein-Zernike equation \cite{msw1,msw2}.
\end {enumerate}

The ion-dipole mixture is the simplest model of a discrete solvent electrolyte which has analytical solutions \cite{blumid}-\cite{lbl16}.
 The solution and thermodynamics are given in terms of three parameters $b_0,b_1,b_2$, which correspond to ion-ion,ion-dipole and dipole-dipole interactions, or their corresponding scaling parameters $\Gamma,\lambda,{\cal B}$, which are functions of only two  coupling parameters for the ions and the dipoles
$
d_0= \kappa_{Debye}$ and $ d_2=3 y_{dipole}
$ (see Eqs.(\ref{d0}),(\ref{d2})).

In this paper we show that introducing two new scaling parameters $\beta_0,\beta_1$ the problem is reduced to only two equations, which are easily solved by iteration. Furthermore, we get a rather compact expression for the excess entropy. The chemical potential of the ions and the dipoles is directly related to the scaling parameters, as we will see below.

This also leads to a natural extension of the Born cycle to arbitrary ionic concentrations, and hence a robust general definition of the Born solvation energy for arbitrary concentrations.\\

The MSA and its extensions have been used in many very interesting applications\cite{nicholls,harvey}. Especially noteworthy are the engineering applications of S.Lvov and his group,\cite{lvov} and the Beijing group\cite{gao,liu}. \\

In section 1 we summarize previous work on the electrostatic equivalent models for ions and for dipoles. In section 2 we review formalism for the ion-dipole mixture using the old scaling parameters $\Gamma,\lambda,{\cal B}$. In section 3 we use a matrix formalism to introduce the new scaling parameters. In section 4 we derive the new results for the thermodynamics and the Born solvation thermodynamics. 

\subsection{ The electrostatic equivalent models}

The MSA can be interpreted using an electrostatic analog from which  all the  results of the MSA can be deriveded \cite{yr6,bvel1,bvel2}. For charges this is simply a spherical capacitor. Here we include also the dipolar case, for which the equivalent capacitor is a dielectric sphere surrounded by a  media of different dielectric constant.

\subsubsection{Charge-Charge Interactions}

 For the primitive model of ionic solutions in the general case \cite{bluma}the
parameter $\Gamma$ is determined from the equation 
\be
 {\frac{d_0^{2}}
{(1+\Gamma  )^2}}= 4\Gamma^{2} \label{eq:n5}
 \ee
\be
d_0^2={ 4 \pi e^2\over{\epsilon_W  k_B T}}\sum_i \rho_i z_i^{2}
 \label{d0}
 \ee

where tho ionic charge is $z_{i} e$ and number density $\rho_{i}%
={\mathcal N}_{i}/V $, where $\mathcal{N}_{i}$ is the number of ions and $V$ is
the volume of the system. Also $\kappa$ is defined by Eq.(\ref{kke}).
We remember that the excess internal energy is
\be
 \Delta E(\Gamma)=-{ 4 \pi e^2\over{\epsilon_W }}\sum_i {\frac{\rho_i z_i^{2}}%
{(1+1/\Gamma  )}} \label{eq:n5n}
 \ee
which is simply the energy density of a collection of spherical capacitors. 

The excess entropy is
\be
 \Delta S^{(MSA)}= - k V \frac{ \Gamma^3}{3 \pi}
\label{ds}
\ee
 Then $\Gamma$ is determined in every case by the simple  variational equation (\ref{A})

\be
\frac{\partial[\beta \Delta E(\Gamma)+\Gamma^3/(3 \pi)]}{\partial \Gamma}=0
\label{clos1}
\ee
The functional form of the entropy is dictated by the excluded volume property of the MSA. Not every functional will guarantee no overlap of the hard cores of the ion pairs.

\subsubsection{Dipole-Dipole Interactions}{dddd}

For a system of hard spheres with a permanent dipole moment
$\mu$ the MSA result can be expressed in terms of a single parameter
$\lambda$ . Following Wertheim\cite{wertheim}, we have

 \be
d_2^{2}=\frac{\lambda^{2}(\lambda+2)^{2}}{9}(1-{\frac{1}{{\epsilon_W}}})
\label{eq:n7c} \ee
where \be
d_2^{2}={\frac{4\pi\rho_s\mu_s^{2}}{{3k_BT}}}=3 y \label{d2} 
\ee
and $\rho_{s}$ is
the solvent number density. y is the adimensional Debye coupling parameter. The MSA dielectric constant
$\epsilon_{W}$  is given by

\be
\epsilon_W=\frac{\lambda^{2}(\lambda+1)^{4}}{16}
 \label{eq:n9}
 \ee

As has been often done in the literature, the parameter $\lambda$ can be
computed directly from the dielectric constant $\epsilon_{W}$ using the above
cubic equation. This parametrization defines an effective polarization
parameter. Just as in the case of the ions, there is a physically meaningful way of interpreting the MSA for point dipoles using the variational principle (\ref{A}). The dipolar system can be represented by a collection  spheres with a dipolar charge distribution \cite{yr5}: The parameter $\lambda$ can be interpreted in terms of a model consisting of a point dipole immersed in a sphere of dielectric constant${\epsilon_{in}}$ which is surrounded by a continuum of dielectric constant ${\epsilon_{out}}
$. We write 

\[
\lambda=\frac{\epsilon_{in}}{\epsilon_{out}}
\]
Now the MSA solution is given by the excess energy parameter $b_2$
\[
\frac{b_{2}}{6}=\frac{\lambda-1}{\lambda+2}=g_{k}^{eff}
\]
where $b_2$ is the dipole-dipole energy parameter defined below Eq.(\ref{eq:w4}). Since this expression is the Clausius-Mossotti equation we can interpret  $g_{k}^{eff}$ as the {\it effective} Kirkwood parameter for this system.
From here we calculate the induced  dipole
\[
{\cal X}_{d}=\frac{3d_{2}}{\lambda+2}=d_{2}\beta_{6}
\]
 since $d_2$ is the dipole of our spheres in reduced units. The excess energy
\be
\beta E=d_{2}{\cal X}_{d}
\label{eeee}
\ee
The closure equation (\ref{eq:n7c}) can be rewritten as
\be
 \frac{9d_2^{2}}
{(\lambda+2)^{2}}=
\lambda^{2}-\frac{16}{(\lambda+1)^{4}}
\ee
This corresponds to exactly Eq.(\ref{A}) in the form

\be
\frac{\partial\beta E}{V\partial\lambda}=-\frac{\pi}{ V k}\frac{\partial
S}{\partial\lambda}=
\lambda^{2}-\frac{16}{(\lambda+1)^{4}}
\ee
which can be integrated to yield\newline 

\be
-\frac{\pi}{ V k} S=\frac{1}{3}\left[
\lambda^{3}+2\left(  \frac{2}{\lambda+1}\right)^{3}\right]  -1
\ee
Now if we define  the scaling lengths for the irrep
$\chi=0$ 
\[\Gamma_{0}=\lambda\]
and for $\chi=\pm1$ \[\Gamma_{1}=\frac
{2}{\lambda+1}.\] 
then \be
 -\frac{\pi}{ V k} S=\frac{1}{3}\left[
\left(  \Gamma_0\right)  ^{3}+2\left( \Gamma_1\right)  ^{3}\right] -1
\ee
Notice that they satisfy the Wertheim 'density' of the
irreps since they are obtained by setting
\[
\rho_{1}=-(1/2)\rho_{0}
\]
Furthermore, the choice 
\[
{\epsilon_{out}}=1\qquad {\epsilon_{in}}=\lambda
\]
which corresponds to placing the dielectric sphere with an embedded dipole in the vacuum satisfies all the physical requirements of the case.
\section{Ion-Dipole Interactions}
\pa{}
In this section we will deduce new equations for the MSA thermodynamics for ion-dipole mixtures. 
\subsection{Summary of previous results}

	For the sake of clarity we summarize former work \cite{blumid,adp,lbl3,harvey,lbl6,lbl7,lbl8}.
Expanding the total pair correlation  $h(12)$  in rotational invariants \cite{blumtorr}
\be 
h(12)= \hat h^{000}(r_{12})+
\hat h^{011}(r_{12})\hat\Phi^{011}+\hat h^{101}(r_{12})\hat\Phi^{101} 
+\hat h^{110}(r_{12})\hat\Phi^{110}+
\hat h^{112}(r_{12})\hat\Phi^{112} 
\label{eq:hexp} 
\ee
 where 
$\hat h^{mn\ell}(r_{12})$ are the coefficient of the invariant expansion, which 
depend only on the distance $r_{12}$ between spheres 1 and 2. The rotational 
invariants $\hat\Phi^{mn\ell}$ depend only on the mutual orientations of the 
 molecules. The relevant correlation functions are 
\begin{itemize} 
\item ion-ion: 
\be 
h_{ii}(r)=(1/2)\left[\hat h^{000}_{++}(r)-\hat h^{000}_{+-}(r)\right] 
\label{eq:w1} 
\ee 
\item
\be 
h_{id}(r)=(1/2)\left[\hat h^{011}_{+d}(r)-\hat h^{011}_{-d}(r)\right]
({\bf{\hat r}} \cdot {\hat{\bf\mu}}) 
 \label{eq:w2} 
\ee 
\item dipole-dipole: 
\[ 
h_{dd}(r)=-\sqrt{3} 
\hat h^{110}_{dd}(r){\hat{\bf\mu}}_1 \cdot {\hat{\bf\mu}}_2 
\] 
\be 
\qquad  +\sqrt{\frac{15}{2}}\hat h^{112}_{dd}(r)
\left[3({\bf{\hat r}} \cdot {\hat{\bf\mu}}_1)({\bf{\hat r}} \cdot 
{\hat{\bf\mu}}_2)-{\hat{\bf\mu}}_1 \cdot {\hat{\bf\mu}}_2\right]  
\label{eq:w3} 
\ee 
\end{itemize}  
where 
${\hat{\bf\mu}}$ is the unit vector in the direction of ${\bf \mu}$. The 
solution of the MSA is given in terms of the 'energy' 
parameters 
\begin{itemize}  
\item ion-ion: 
\be b_0=2 \pi \rho_i 
\int_{0}^{\infty} dr h_{ii}^{000}(r)r  
\label{eq:w3a} 
\ee 
\item 
ion-dipole:  
\be b_1=2 \pi \sqrt{\frac{\rho_i  \rho_s}{3}}\int_{0}^{\infty}dr 
h_{id}^{101}(r)   
\label{eq:w4} 
\ee 
\item dipole-dipole: 
\be b_2=3 \pi  \rho_s 
\sqrt{\frac{2}{15}} \int_{0}^{\infty} dr \frac{\hat h^{112}(r)}{r} 
\label{eq:w5} 
\ee         
\end{itemize} 
which, as will be shown below are proportional to the ion-ion, ion-dipole and dipole-dipole excess 
internal energy\cite{lbl3}. These parameters are required to satisfy the following equations 
\cite{blumid} 
 
\be 
a_1^2+a_2^2=d_0^2 
\label{eq:w6a} 
\ee 
\be 
a_1 K_{10} - a_2[1-K_{11}]=d_0d_2
\label{eq:w7a} 
\ee 
\be 
 K_{10}^2 +[(1-K_{11})]^2=y_1^2+ d_2^2
\label{eq:w8a} 
\ee

where $d_0$ and $d_2$ are defined by Eqs.(\ref{d0})and (\ref{d2}). Also
\be 
{\cal D}= 1+{\cal B} 
\label{eq:v3} 
\ee
with
\be
{\cal B} =\frac{b_1^2}{4\beta_6^2}=\frac{b_1^2(\lambda+2)^2}{36}
\label{eq:v2}
\ee
\[
\beta_6=1-\frac{b_2}{6}
\]
 In our previous work simple sets of equations were obtained using 
the  scaling lengths \cite{lbl6,lbl7} $\Gamma,\lambda$ and $\cal B$. 
These are related to the excess energy parameters $b_0,b_1$ and $b_2$ of  Eqs.(\ref{eq:w3a}-\ref{eq:w5}):
\be 
b_0=\frac{-\Gamma }{1+\Gamma }+{\cal B} \left[\frac{1}{1+\Gamma}+\frac{1}{2+\lambda}
\right]\label{w15} 
\ee
\be
b_1=\sqrt{{\cal B}}\frac{6}{2+\lambda}
\label{w16}
\ee
and
\be
b_2=6\frac{\lambda-1}{2+\lambda}
\label{w17}
\ee
Then the closure equations {\ref{eq:w6a}-\ref{eq:w8a}} are:
 \begin{itemize}
\item Ion-ion
\be 
 \frac{d_0}
    {( 1 + \Gamma  ) \sqrt{\epsilon_w}}
     =2\Gamma  {\cal D}^2  \left[1+  \frac{
          \left\{  
         {d_0}{\sqrt{\epsilon }_w}/( 2 + 2\Gamma )-\Gamma\right\}}{
         \left( 1 + \lambda  \right) \,
         \left( \Gamma  + \lambda  \right) } \right]
\label{eq:w6b} 
\ee 
\item Dipole-dipole

\be
\frac{ (y_1^2+ d_2^2)}{( 2 + \lambda )^2 }
=\frac{1}{9\,\cal D }\,
    \left( {\lambda }^2 + 
      \frac{{\cal B} {\cal F} }{( 2 + 
            \lambda )^2}
          \right) 
\label{eq:w8b} 
\ee
with
\be
{\cal F}= {( 1 + \Gamma  + \lambda  ) }^2\,
            {( 3 + \Gamma  + \lambda  ) }^2 - 
           \frac{1}{\cal D }{( 1 + \Gamma  ) }^2
              ( 1 + \lambda  ) \,
              ( 5 + 2\Gamma  + 3\lambda  ) 
\label{calF} 
\ee
\noindent and
\item Ion-dipole: This equation is in its original form

\[ 
\frac{d_0 d_2}{( 1 + \Gamma  )}=
\]
\be
\frac{2\sqrt{\cal B} }{3\cal D }\left[ {\left( 1 + \Gamma  + \lambda  \right) }^2\,
       \left( 3 + \Gamma  + \lambda  \right)  - 
      \frac{\left( 1 + \Gamma  \right) \,
         \left( 1 + \lambda  \right) \,
         \left( 3 + 2\,\Gamma  + 2\,\lambda  \right) }{
         \cal D } \right] 
\label{eq:w7b} 
\ee 
which is really equivalent to the linear  explicit equation for ${\cal B}$ \cite{bf2}
\be
{\cal B}=\frac{\Gamma \,\lambda }{1 + 
    \Gamma  + \lambda }\left[\frac{-8\,\Gamma \,\left( 1 + \Gamma  \right)  + 
  {d_0}\,\lambda \,
   {\left( 1 + \lambda  \right) }^2}{ {d_0}\,\lambda \,
   {\left( 1 + \lambda  \right) }^2 + 
  8\,\left( 1 + \Gamma  \right) \,\lambda \,
   \left( 1 + \Gamma  + \lambda  \right)}\right]
\ee
so that our problem is reduced to  a set of two equations for the two scaling parameters $\Gamma, \lambda$.
\end{itemize}

Using Eq(\ref{eq:w6b}) we write
\be
{\cal B}= \frac{{\Gamma }^2\,\lambda \,
    {{\epsilon }_w}}{
     {\cal D} \,
    \left( 1 + \Gamma  + 
      \lambda  \right) }
{\cal Y}
\ee
with
\be
{\cal Y}=
\left[
1 - \frac{{\cal B} \,
     \left( 1 + \Gamma 
       \right) \,
     \left( 1 + \lambda 
       \right) }{
      {\cal D} \,
     \Gamma \,\lambda } - 
  \frac{{\cal D} }
   {{{\epsilon }_w}}
\right]\left[
\frac{1}
  {\lambda \,
     \left( 1 + \Gamma  + 
       \lambda  \right)  + 
    \frac{{d_0}\,
       {\sqrt{{{\epsilon }_
          w}}}}{2\,
       \left( 1 + \Gamma 
         \right) }}\right]
\label{yyyy}
\ee
For infinite dilution we get the limit
\[
{\cal B}=
\frac{\Gamma^2}{( 1 + \lambda 
        )^2}
    \left[    
       {\epsilon _w}-1
      \right] 
\]
The interesting remark is that using the definition of ${\cal B}$, Eq.(\ref{eq:v2}), we get for the ion-dipole excess energy parameter $b_1$ Eq.(\ref{eq:w4})
\be
b_1=
d_0 
    \frac{3}
    {( 1 + \lambda )( 2 + \lambda ) }
   \sqrt{ \left( 1 - 
      \frac{1}
       {\epsilon_w}
      \right)} 
\label{bornca}
\ee
An efficient iterative solution of the equations is to start assuming that ${\cal B}=0$, and compute $\lambda$ 

\begin{enumerate}
\item Then solve Eq.(\ref{eq:w8b}) by iteration. Since $\lambda>1$ 
\be
\lambda=-1 + {\sqrt{1 + \frac{3 d_2}{\cal L}}}
\ee
with
\be
{\cal L}=\sqrt{1- \frac{1}{\epsilon_w} + \frac{{\cal B}} {{\cal D} }
        \left[ -1 + \frac{{\cal F} }
           {{\lambda }^2\,
             {\left( 2 + \lambda  \right) }^2} \right] }        
\label{lmb}
\ee
where $\cal F$ is defined by Eq.(\ref{calF}).

\item  We get the physical branch of the ion-ion interaction equation (\ref{eq:w6b})
\be
\Gamma=\frac{1}{2}\left[-1 + \sqrt{1 + 2\frac{ d_0}{\sqrt{\varepsilon_w}{\cal G}}}\right]
\ee
\be
{\cal G}={\cal D }^2\left[ 1 + 
    \frac{-2\Gamma ( 1 + \Gamma  )  + 
       d_0
        \sqrt{\epsilon_w} }{2\,
       ( 1 + \Gamma  )
       ( 1 + \lambda  ) 
       ( \Gamma  + \lambda ) } \right]
\label{eq:w6c} 
\ee 
\end{enumerate}

 We illustrate  the results of this procedure with a numerical example in which the solvent is 'dipolar' water of dielectric constant $\epsilon_w=78.4$, density $\rho=0.0332$ and diameter
$\sigma_w=2.76 \AA$.

\begin{figure}
\centereps{10cm}{7cm}{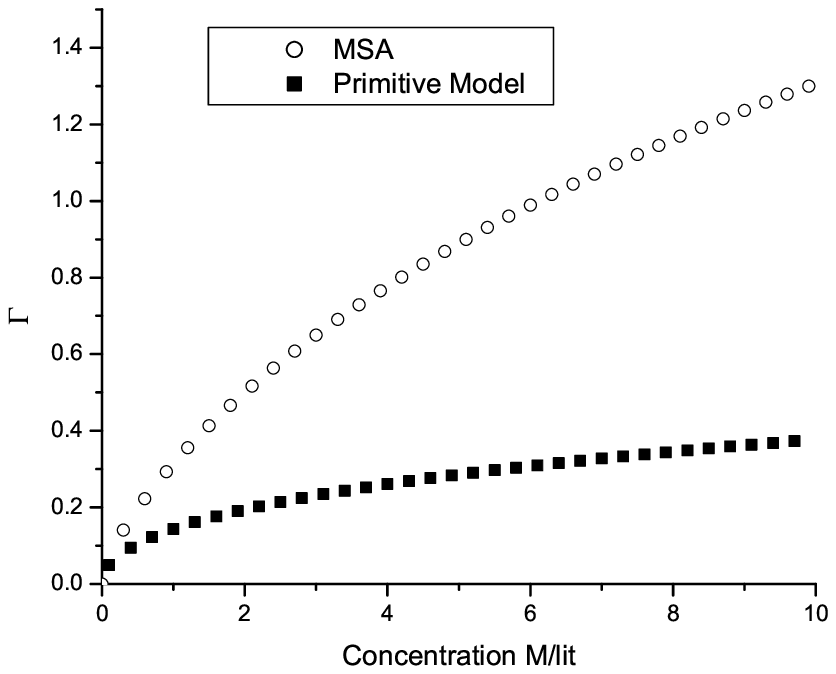}
\parbox{4.5in}{
\caption{ The screening MSA parameter $\Gamma$ for the ion-dipole mixture compared to the primitive model $\Gamma_0$ which assumes a  dielectric constant $\epsilon_w=78.4$}}
\label{fig:one}
\end{figure}

\begin{figure}
\centereps{10cm}{7cm}{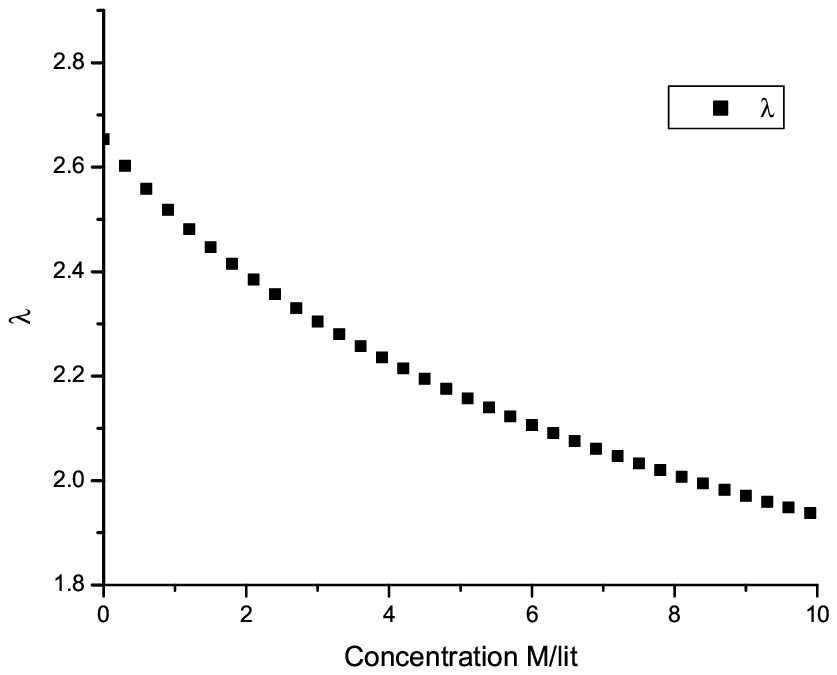}
\parbox{4.5in}{
\caption { The polarization MSA parameter $\lambda$ for the ion-dipole mixture 
as a function of the molar concentration c}}
\label{fig:two}
\end{figure}
  The dipole moment of the 'water' is $\mu=2.21 $ Debyes, which is in ther range of pretty much any other model of water.For the time being we discuss only the restricted primitive model where everybody has the same diameter, which ideally would correspond to a solution of NaF. The general case of arbitrary sizes is left to  future publications. 

\begin{figure}
\centereps{10cm}{7cm}{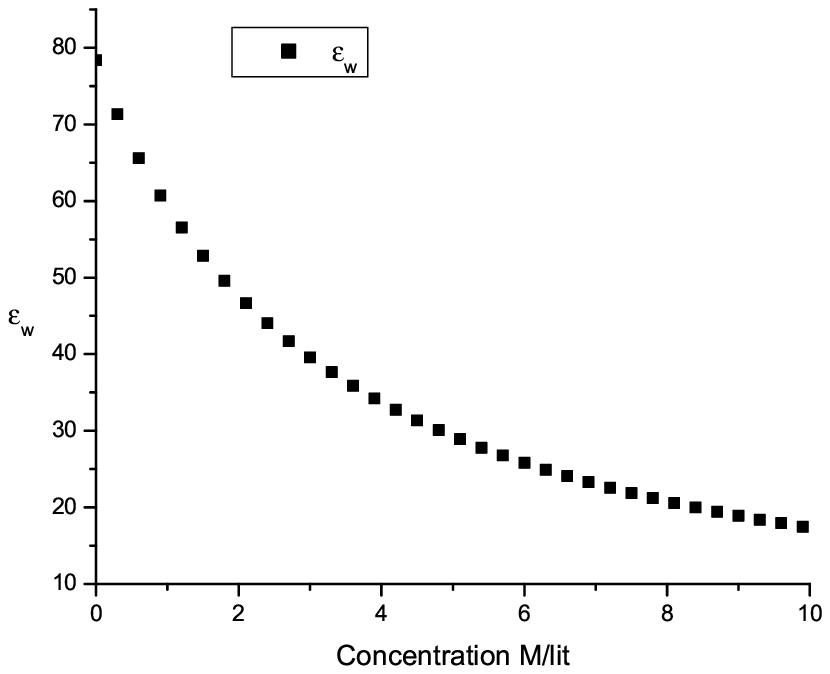}
\parbox{4.in}{
\caption{ The dielectric constant $\epsilon_w$ as a function of the molar concentration c}
}
\label{fig:three}
\end{figure}
\begin{figure}
\centereps{10cm}{7cm}{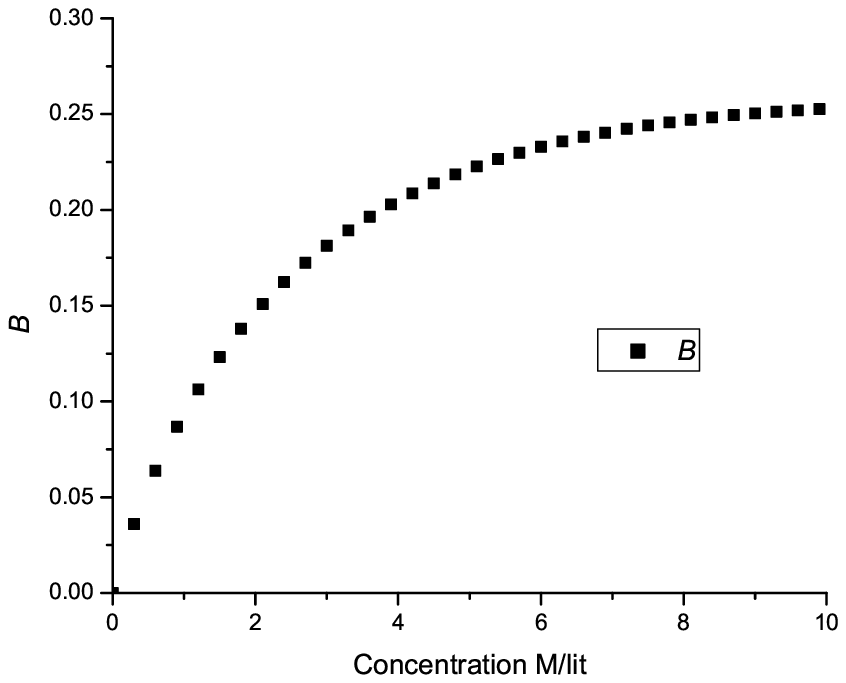}
\parbox{4.in}{
\caption{ The reduced ion-dipole MSA parameter $\cal B$ as a function of the molar concentration c}
}
\label{fig:four}
\end{figure}
In Fig. 1 to 4 we show the results for the MSA parameters $\Gamma,\lambda,\cal B$, and the dielectric constant $\epsilon_w $ as a function of  molar concentration of a 1-1 salt for the restricted primitive model in which the diameters are all equal to 1.

\subsubsection{ The scaled relations}

Using the scaling results   
 \cite{blum2,lbl3,lbl6,lbl7,lbl8} we find the new set of relations \\

\[
Q_{ii}^0=2-\frac{\beta_6}{D_f}=2\left(1-\frac{1+\Gamma}{{\cal D}}\right)=2\left(\frac{1}{{\cal D}}\right)\{\Gamma+{\cal B}\}
 \]
\[
Q_{id}^0=\frac{ b_1}{2 D_f}=\left(1-\frac{1+\Gamma}{{\cal D}}\right)\sqrt{{\cal B}}=\left(\frac{1}{{\cal D}}\right)\{\Gamma+{\cal B}\}\sqrt{{\cal B}}
\]
\be
Q_{dd}^0=2-3\frac{1+b_0}{ 2 D_f}=\left(\frac{1 
}{\cal D}\right)\{-\lambda+{\cal B}(1+\Gamma +\lambda)\}
\ee
\be
Q'_{ii}=\left(\frac{2}{\cal D}\right)\{\Gamma^2
+{\cal B}\}
\label{qpii}
\ee

\be
Q'_{id}=\left(\frac{2 \sqrt{{\cal B}}
}{\cal D}\right)\{1+\Gamma\}\{\Gamma +\lambda\}
\label{qpid}
\ee

\be
Q'_{dd}=\left(\frac{2 
}{\cal D}\right)\{1-\lambda^2+{\cal B}(2+\Gamma +\lambda)\}
\label{qpdd}
\ee

with
\be
D_f=\frac{\beta_6}{2 (1+\Gamma \sigma)} \left[1+\left(\frac{b_1 (2+\lambda)}{6}\right)^2\right]=\frac{\beta_6 {\cal D}}{2 (1+\Gamma \sigma)}
\ee
\be
{\cal B}=\left(\frac{b_1 (2+\lambda)}{6}\right)^2
\ee
 Also
\be
q'=2-8\frac{1}{(1+\lambda)^2}
\ee

\section{  Matrix Reformulation}

To obtain further simplifications  we  need to rewrite the  solution of the ion dipole MSA \cite{blum2,lbl3} in matrix form. Remember that
\be
{\bf Q}(r)=(r-1){\bf Q}'+(1/2)(r-1)^2 {\bf Q}"-{\bf A}
\ee
so that
\be
{\bf Q}(0)=-{\bf Q}'+(1/2) {\bf Q}"-{\bf A};\qquad
{\bf Q}'(1)={\bf Q}';\qquad 
{\bf Q}'(0)={\bf Q}'- {\bf Q}"
\ee
with
\begin{equation}
{\bf Q }'=\left[ 
\begin{array}{cc}
Q'_{ii} & Q'_{id}\\ 
Q'_{di} & Q'_{dd}  
\end{array}
\right]
\end{equation}
\begin{equation}
{\bf Q }"=\left[ 
\begin{array}{cc}
0 & 0\\ 
Q"_{di} & Q"_{dd}  
\end{array}
\right]
\end{equation}

\[
{\bf A}=\left[ 
\begin{array}{cc}
a_1 & a_2\\ 
0 & 0  
\end{array}
\right]
\]

 The factor function matrix at the origin is
\begin{equation}
{\bf Q }(0)=\left[ 
\begin{array}{cc}
-Q'_{ii}-a_1 & -Q'_{id}-a_2\\ 
-Q'_{di}+Q"_{di}/2 & -Q'_{dd}+Q"_{dd}/2  
\end{array}
\right]
\end{equation}

We define first the matrix
\begin{equation}
{\bf D }_f=\left[ 
\begin{array}{cc}
1+b_0 & b_1/6\\ 
b_1/6 & \beta_6/3  
\end{array}
\right]
\end{equation}
with the inverse
\begin{equation}
{\bf D }_f^{-1}=\left[\frac{3}{2 D_f}\right]\left[ 
\begin{array}{cc}
\beta_6/3 & -b_1/6\\ 
-b_1/6 &1+b_0   
\end{array}
\right]
\end{equation}

We get 
\be
{\bf A}=-2\left[ 
\begin{array}{cc}
1& 0 \\ 
0&0\end{array}
\right]\cdot{\bf D}_f^{-1}\cdot \left[  {\bf I}_2-{\bf D}_f^{-1}\right]
\ee
where
\[ {\bf I}_2=\left[ 
\begin{array}{cc}
1& 0 \\ 
0&2\end{array}
\right]
\]

and
\be
{\bf Q}(0)=2 {\bf I}-{\bf D }_f^{-1}=2{\bf Q}^0+
\left[ 
\begin{array}{cc}
0 &0\\ 
0 & 2  
\end{array}
\right]
\label{q0d}
\ee
where ${\bf I}$ is the 2 dimensional unit matrix and 
\be
{\bf Q}^0=
\left[ 
\begin{array}{cc}
Q^0_{ii} &\frac{ Q^0_{id}}{2}\\ 
\frac{ Q^0_{id}}{2} & Q^0_{dd}  
\end{array}
\right]
=\left[ 
\begin{array}{cc}
1-\frac{\beta_6}{2 D_f} & \frac{b_1}{4 D_f}\\ 
\frac{b_1}{4 D_f}&2-\frac{3(1+b_0)}{2 D_f } 
\end{array}
\right]
\end{equation}

 We use the well known relation
\be
{\bf Q}'(0)+\left[{\bf Q}'(0)\right]^T=-{\bf Q}(0)\cdot{\bf Q}(0)
\ee
to find the matrix ${\bf Q}'(1)$ (we use $\sigma=1$ throughout) which is also the 
matrix of the contact pair correlation functions:
\be
{\bf Q}'(1)=\left[ 
\begin{array}{cc}
-2 &0\\ 
0 & 2  
\end{array}
\right]
\cdot
\left[ 
\begin{array}{cc}
Q^0_{ii} &\frac{ Q^0_{id}}{2}\\ 
\frac{ Q^0_{id}}{2} & Q^0_{dd}  
\end{array}
\right]\cdot
\left[ 
\begin{array}{cc}
Q^0_{ii} &\frac{ Q^0_{id}}{2}\\ 
\frac{ Q^0_{id}}{2} & Q^0_{dd}  
\end{array}
\right]-
\left[ 
\begin{array}{cc}
0& 0\\ 
0&2
\end{array}
\right]
\label{Qpr}
\ee

This immediately suggests the new scaling parameters $\beta_0,\beta_1$
\be
\stackrel {\rightarrow}{\beta}=
\left[ 
\begin{array}{c}
-\frac{\beta_0}{2}\\ 
\beta_1
\end{array}
\right]
\ee
with
\[
\beta_0=(1+b_0)d_0-b_1d_2
\]
\[
\beta_1=-\frac{b_1 d_0}{12}+\beta_6 d_2
\]
 Then, we get
\be
{\bf Q}'(1)=\left[ 
\begin{array}{cc}
-1/2\beta_0^2 &\beta_0 \beta_1 \\
-\beta_0 \beta_1 &-2+  2 \beta_1 ^2 
\end{array}
\right]
+y_1^2\left[ 
\begin{array}{cc}
-b1^2/2&- b_1 \beta_6\\ 
 b_1 \beta_6 &2\beta_6^2
\end{array}
\right]
\end{equation}

From Eq.(\ref{Qpr}) we see that
\[
\beta_0=
\left[ -b_1^2 y_1^2+\frac{1}{D_f^2}[b_1^2/4+(b_0 \beta_6-b_1^2/12)^2]\right]^{1/2}
\]
\be
\qquad=2 {\sqrt{\frac{\Gamma ^2}{{\cal D }}+ {\cal B } \left( \frac{1}{\cal D } - \frac{16}{{\left( 
                          1 + \lambda  \right) }^4} \right) }}
\label{beta0}
\ee
and 
\[
\beta_1=
\left[ -\beta_6^2 y_1^2+\frac{1}{D_f^2}[b_1^2/16+((1+b_0) \beta_3/2+b_1^2/12)^2]\right]^{1/2}
\]
\be
\qquad={\sqrt{{\lambda}^2 - \frac{16} {{\left(1 + \lambda \right)}^4} +
        \frac{{
            \cal B}\left(1 + \Gamma \right)\,
            \left(1 + \Gamma + 2\, \lambda \right)}{\cal D }}}
\label{beta1}
\ee
 Furthermore, we get the new cross relation
\be
\left[b_1 \beta_0/4-\beta_1 \beta_6\right]^2=b_1^2/4+\beta_3^2-y_1^2 \Delta^2
\ee
 We observe also that the closure relations Eqs(\ref{eq:w6a}-\ref{eq:w8a}) are equivalent to

\be 
a_1^2+a_2^2=d_0^2 
\label{eq:w6} 
\ee 
\be 
\frac{a_1 b_1}{2}-a_2\beta_3=d_0  
\Delta {\cal A} 
\label{eq:w7} 
\ee 
\be 
\left[\frac{ b_1}{2}\right]^2+\beta_3^2= \Delta^2
{\cal A}^2 +y_1^2 \Delta^2
\label{eq:w8} 
\ee 
where we have used the definition
\be
{\cal A}=d_2-\frac{b_1 d_0}{4\Delta 
}\left[1+ b_0+\frac{\beta_6 }{3}\right] =\frac{\beta_6}{\Delta}\left[ \frac{\beta_0}{2}\sqrt{{\cal B}}-\beta_1 \right]
 \label{eq:w13} 
\ee

\section{ Thermodynamics}
We start with the equations derived  by Vericat and Blum \cite{lbl3,hs1}: The internal energy is  
\be
\beta E/V=
\frac{1} {4\pi }\left[ b_0
       {d_0}^2 -  2 b_1 d_0\,  d_2-2 b_2 d_2^2\right] 
\label{EE}
\ee
which can be written as 
\be
\beta E/V=
\frac{1} {4\pi }\left[ 
       d_0 (\beta_0-d_0)+ 12 d_2 (\beta_1-d_2)\right] 
\label{EEb}
\ee

For the Helmholtz free energy we got
\be
\beta A/V=-
\frac{1} {12\pi }\left( -2\,b_0\,
       {d_0}^2 +  2 b_1 d_0\,  d_2 + 
      2\,{q'}^2 + {Q'_{dd}}^2 +  2\,{Q'_{id}}^2 + 
     [ Q'_{ii}]^2 \right) 
\label{AA}
\ee

\begin{figure}
\centereps{10cm}{7cm}{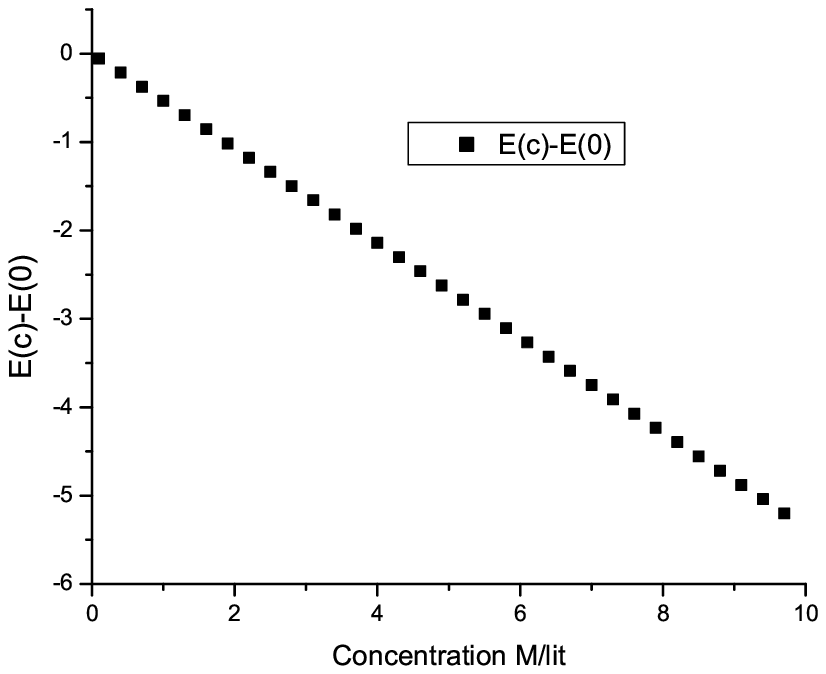}
\parbox{4.5in}{
\caption{ The excess internal energy $\beta(E(c)-E(0))/V$ as a function of the molar concentration c.}
}
\label{fig:five}
\end{figure}
From here the entropy can be computed using 
\be
\frac{S}{k V}=\frac{\beta}{V}[E-A]
\ee
which leads to
\be
\frac{S}{k V}=\frac{1} {12\pi }\left( b_0  {d_0}^2 - 4\ b_1
d_0 d_2 - 6 b_2 d_2^2+ 2\,{q'}^2 + {Q'_{dd}}^2 +  2\,{Q'_{id}}^2 + [Q'_{ii}]^2 \right) 
\ee
 
Using the results of last section we get 
\[
S= \frac{k V}
    {3\pi }
\]
\be
\left[{3 + \frac{16\,\cal D }
     {{\left( 1 + \lambda  \right) }^4} - 
    \frac{32}{{\left( 1 + \lambda  \right) }^3} - 
     {\Delta_ \beta}\,
     {\sqrt{\frac{1}{4} + { {\Delta _\beta}}^4 + 
         \frac{16(\lambda -1) }
          {{\left( 1 + \lambda  \right) }^3} - 
         {\left( \frac{1}{2} - 
             \frac{16\,\cal D }
              {{\left( 1 + \lambda  \right) }^4} \right) }^2}}}\right]
\ee
where
\be
\Delta_\beta=\sqrt{\beta_0^2/4+\beta_1^2}
\label{delta}
\ee
\begin{figure}
\centereps{10cm}{7cm}{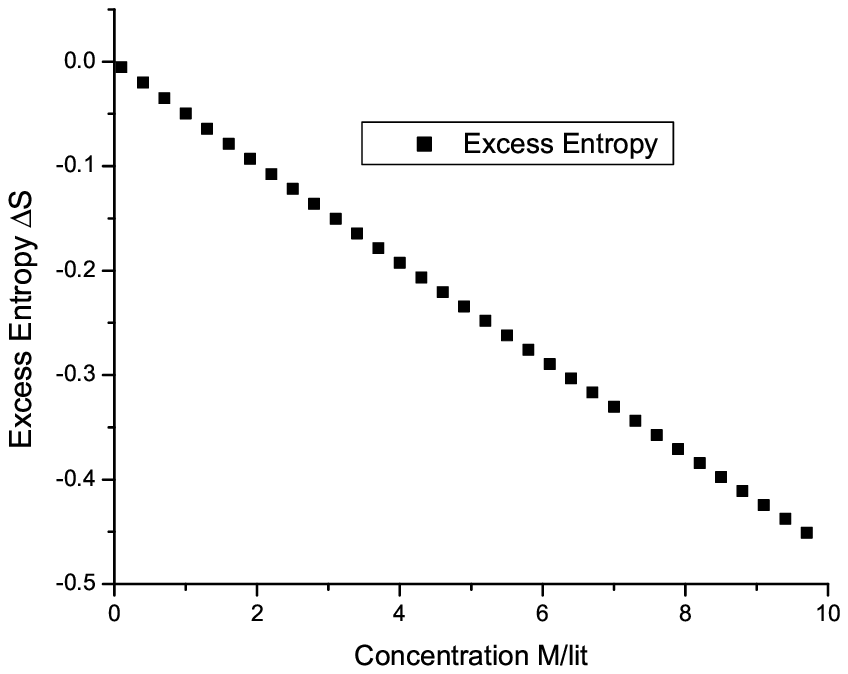}
\parbox{4.5in}{
\caption{ The excess entropy $\beta(S(c)-S(0))/V$ as a function of the molar concentration c.}
}
\label{fig:six}
\end{figure}

The ionic excess energy as a function of the molar concentration is shown in Fig. 5. The excess entropy is shown in Fig. 6.

This result can also be expressed in terms of the regular scaling parameters $\Gamma,\lambda,{\cal B}$ \cite{lbl6}

\be
S= \frac{k V}
    {3\pi }
\left[
{3 + \frac{16\,\cal D }
     {{\left( 1 + \lambda  \right) }^4} - 
    \frac{32}{{\left( 1 + \lambda  \right) }^3} - 
     {\Gamma_T}\,
     {\sqrt{\frac{1}{4} + 
         \frac{16\,\left( -1 + \lambda  \right) }
          {{\left( 1 + \lambda  \right) }^3} - 
         {\left( \frac{1}{2} - 
             \frac{16\,\cal D }
              {{\left( 1 + \lambda  \right) }^4} \right) }^2 + 
         \Gamma_{T}^4}}}\right]
\ee
with
\be
\Gamma_T=
{\sqrt{{\Gamma }^2 + {\lambda }^2 - 
     \frac{16\,\cal D }
      {{\left( 1 + \lambda  \right) }^4} + 
     \frac{2\,{\cal B} \,\left( 1 + \Gamma  \right) \,
        \left( 1 + \lambda  \right) }{\cal D }}}
\label{skv8n}
\ee

\subsection{The activity and osmotic coefficients: An extended Born solvation formula}
Using the expression of the Helmholtz free energy Eq.(\ref{AA}) we derive the chemical potentials of the ions
\be
\beta \mu_i=\frac{(\partial \beta A/V)}{\partial \rho_i}
\label{mui}=\frac{d_0^2}{4 \pi \rho_i}\left[\frac{ \beta_0}{ d_0}-1\right]=\frac{\beta e^2}{\sigma_i }\left[\frac{ \beta_0}{ d_0}-1\right]
\ee
and of the dipoles
\be
\beta \mu_d=\frac{(\partial \beta A/V)}{\partial \rho_d}
\label{mud}=\frac{12 d_2^2}{4 \pi \rho_d}\left[\frac{ \beta_1}{ d_2}-1\right]=\frac{4\beta \mu^2}{\sigma^3}\left[\frac{ \beta_1}{ d_2}-1\right]
\ee
 Then the Gibbs free energy is
\be
\frac{\beta G}{V}= \rho_i \beta \mu_i+\rho_d \beta \mu_d=\frac{\beta E}{V}
\label{GG}
\ee
where we used Eq.(\ref{EEb}).\\

 The new scaling parameters 'diagonalize' the internal  energy, since the internal energy consists now of only two terms
\be
E/V=\frac{ e^2}{\sigma_i }\left[\frac{ \beta_0}{ d_0}-1\right]
+ \frac{4 \mu^2}{\sigma^3_d}\left[\frac{ \beta_1}{ d_2}-1\right]=\frac{1}{V}[\Delta E_{ionic}+\Delta E_{dipole}]
\label{EEborn}    
\ee

The first term clearly is the charge energy term and the second term is the dipolar contribution, which however is dependent on the ion concentration. Since they are strictly additive one can construct a Born cycle for any arbitrary concentration.

From Eq.(\ref{beta0}) we get a general the Born solvation energy \cite{born}.
\be
E_{Born}=\mu_i=\left(\frac{e^2}{\sigma_i}\right)\left[\frac{\Gamma \lambda \,{\sqrt{\frac{{\cal D}}{\epsilon _w} \,
         \left( 1 + 
           {\cal Y}  
           \right) }}}{
    \left( 1 + \Gamma  \right) \,
    \left( \Gamma \,\lambda  - \frac{{\cal B} \,
         \left( 1 + \Gamma  \right) \,
         \left( 1 + \lambda  \right) }{{\cal D} \,
         } \right) }
-1\right]
\label{mub}
\ee
where ${\cal Y}$ is defined by Eq.(\ref{yyyy}). The  dilute limit expressions for  $b_1$, the  ion-dipole excess energy parameter were given by Blum, Chan et al and Zhou et al  \cite{lbl2,chan,zstell}. In all of these the solvent polarization contribution   discussed by Kusalik 
and Patey \cite{kusalik} is missing. Our current definition includes this contribution. 
 Taking the zero concentration limit of Eq.(\ref{mub}) we 
get the limiting Born solvation energy for a molecular solvent, which agrees with previous results as can be seen from equations (\ref{eq:n7c}) and ( \ref{eq:w4}).
We get
\be
\beta \mu_i(0)=-\left(\frac{560.0}{\sigma_i \AA}\right)\left[\frac{\lambda_0}{1+\lambda_0}\right]\left\{1-\frac{1}{\epsilon_{w,0}}\right\}
\ee
where the subindex 0 indicates the pure solvent parameters. 
A simple approximation is obtained when the concentration dependent $\lambda$ and $\epsilon_w$ are used:
\be
\beta  E^{approx}_{Born}=\beta \mu_i(\rho_i)=-\left(\frac{560.0}{\sigma_i \AA}\right)\left[\frac{\lambda(\rho_i)}{1+\lambda(\rho_i)}\right]\left\{1-\frac{1}{\epsilon_{w}(\rho_i)}\right\}
\label{disbornc}
\ee
\newpage
We compare in Fig. 7 the exact expression and the approximate low density one, but with the correct density dependent $\lambda$ and $\varepsilon_w$.\\

\begin{figure}
\centereps{10cm}{7cm}{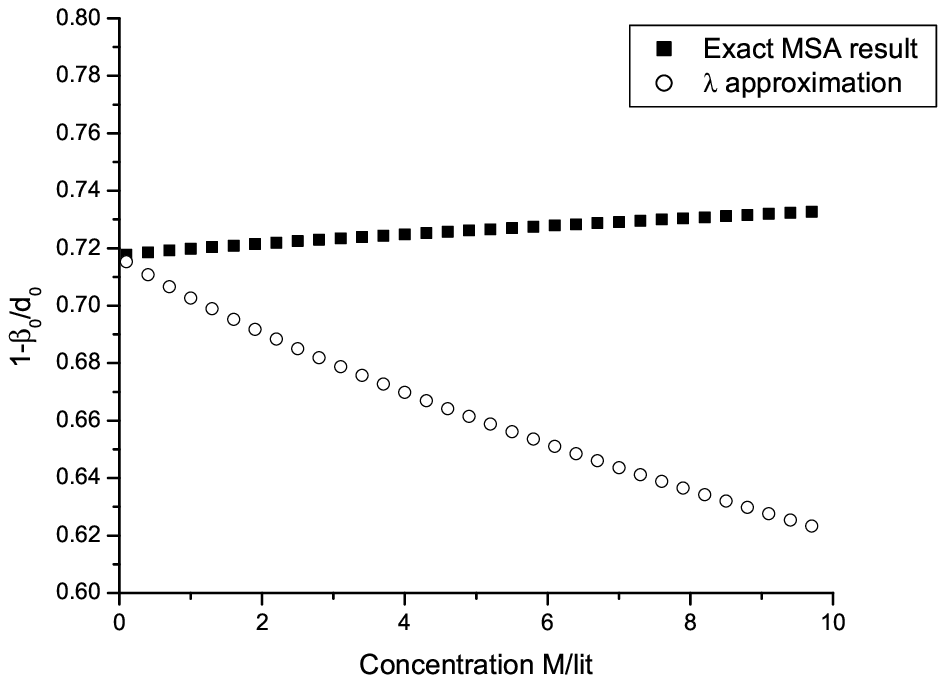}
\parbox{4.5in}{
\caption{ Comparison of the MSA exact result for the Born solvation energy Eq.(\ref{mub}) to  the $\lambda$ approximation  Eq.(\ref{disbornc})}}
\label{fig:seven}
\end{figure}

The activity coefficient of the salt referred to infinite dilution is

\begin{figure}
\centereps{10cm}{7cm}{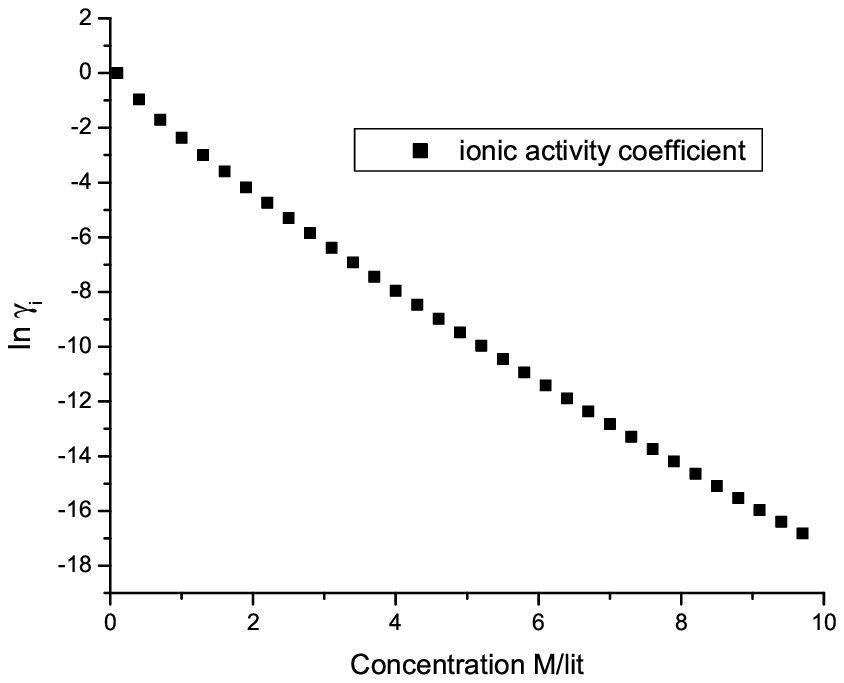}
\parbox{4.5in}{
\caption{ The ionic activity coefficient $ log{\gamma_\pm}$ as  a function of the molar concentration}}
\label{fig:eight}
\end{figure}

\be
\ln {\gamma_\pm}=\left(\frac{560.0}{\sigma_i \AA}\right)\left[\frac{\Gamma \,\lambda {\sqrt{\frac{{\cal D}}{\epsilon _w} \,
         \left( 1 + 
           {\cal Y}  
           \right) }}}{
    \left( 1 + \Gamma  \right) \,
    \left( \Gamma \,\lambda  - \frac{{\cal B} \,
         \left( 1 + \Gamma  \right) \,
         \left( 1 + \lambda  \right) }{{\cal D} \,
         } \right) }
-\left[\frac{1}{1+\lambda_0}\right]\left\{1+\frac{\lambda_0}{\epsilon_{w,0}}\right\}
\right]
\label{ionact}
\ee

Fig. 8 shows the ionic activity coefficient as  a function of the molar concentration.

Similarly, we get the corresponding expressions for the solvent  using Eq.(\ref{beta1})

\[
\beta \mu_d= \frac{\partial (\beta A/V)}{\partial \rho_d}
\]
\be
\qquad= \frac{ 3d_2^2}{ \pi \rho_d}\left[\frac{3}
    {2 + \lambda }\,{\sqrt{\frac{\frac{{\cal B}}
             {{\lambda }^2} \,
             \left( 1 + \Gamma  \right) \,
             \left( 1 + \Gamma  + 2\,\lambda  \right)  + 
          {\cal D} \,
           \left( 1 - \frac{1}{{{\epsilon }_w}} \right) }
          {1 + \frac{9\,{\cal B} \,
             {\cal H} }{{\lambda }^2\,
             {\left( 2 + \lambda  \right) }^2} - 
          \frac{{\cal D} }{{{\epsilon }_w}}}}}-1\right]
\label{beta1d2}
\ee

with
\[
{\cal H}=\frac{{\left( 1 + \Gamma  + \lambda  \right) }^2\,
     {\left( 3 + \Gamma  + \lambda  \right) }^2}{9} - 
  \frac{{\left( 1 + \Gamma  \right) }^2\,
     {\left( 1 + \lambda  \right) }^2\,
     \left( 3 + \frac{2\,\left( 1 + \Gamma  \right) }
        {1 + \lambda } \right) }{9\,{\cal D} }
\]
\begin{figure}
\centereps{10cm}{7cm}{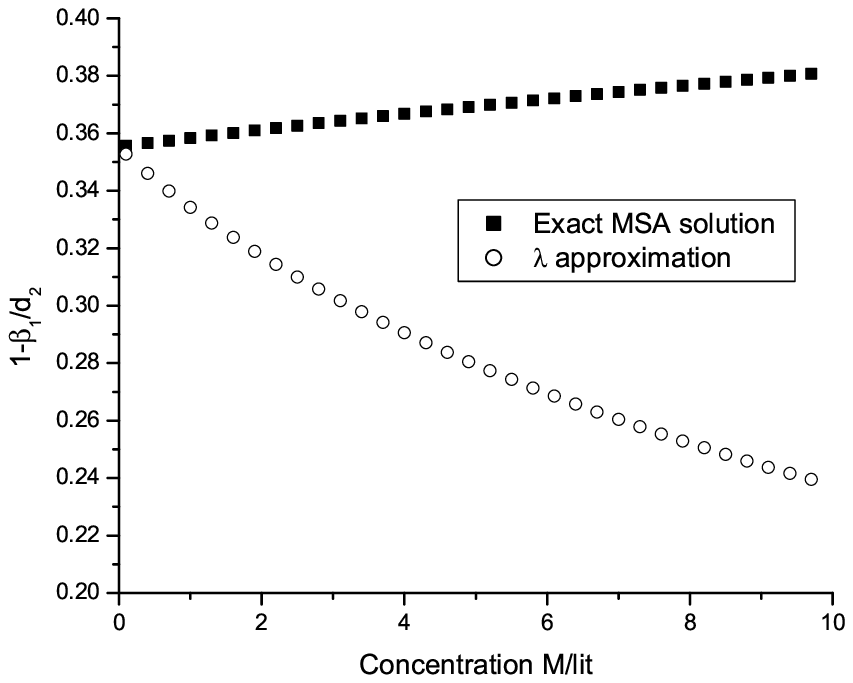}
\parbox{4.5in}{
\caption{ Comparison of the MSA exact result for dipolar   solvation energy Eq.(\ref{beta1d2}) to  the $\lambda$ approximation.)}}
\label{fig:nine}
\end{figure}
 and the zero concentration limit is
\be
\beta \mu_d(0)=  \frac{ \mu^2}{ k T}\left[\frac{3}
    {2 + \lambda_0 }\right]= 865.8 erg/mol\left[\frac{3}
    {2 + \lambda_0 }\right];\quad T=298.13 ^0 K
\ee
which corresponds  to the polarization energy  Eq.(\ref{eeee})discussed in section \ref{dddd}.

A simple approximation for highly polar solvents like water is to  use the full concentration dependent  $\lambda$ in these expressions. Comparisons are shown in Fig.9 and 10 for the ion hydration  and solvent hydration fromulas.

\begin{figure}
\centereps{10cm}{7cm}{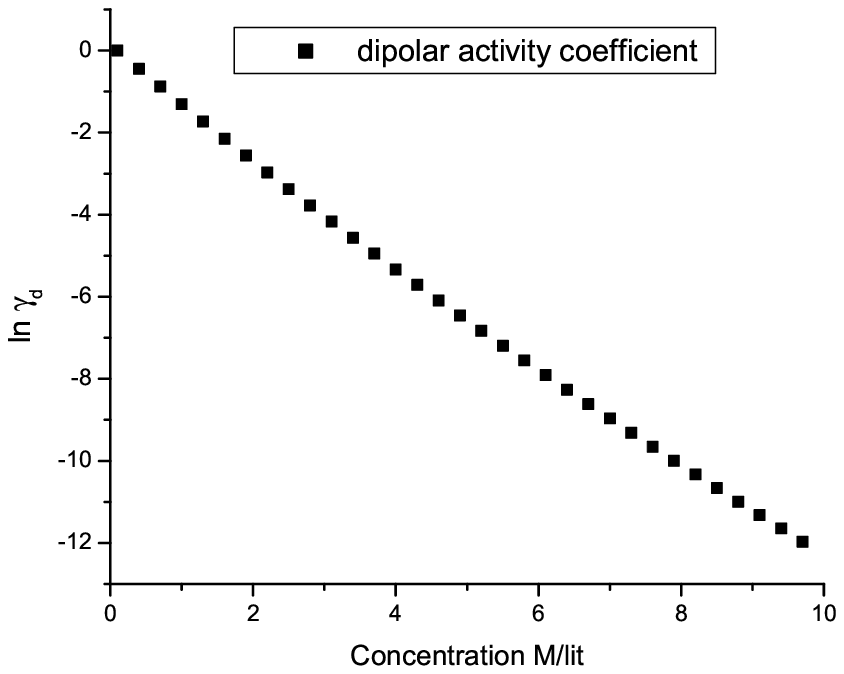}
\parbox{4.5in}{
\caption{  The dipolar activity coefficient $ log{\gamma_d}$ as  a function of the molar concentration}}
\label{fig:ten}
\end{figure}

We remark that in the MSA  the relation,
\be
G= E
\label{eq:c6z2} 
\ee
 still holds, and our chemical potentials satisfy the Gibbs-Duhem relation.

 The excess pressure can also be computed \cite{hs1}. The expression is
\cite{lbl6} 
\be
P/k_BT= S/Vk_BT 
\label{eq:c6z} 
\ee 

The extensions to the general arbitrary size mixtures, as well as applications will be discussed in a forthcoming publication. 
\pa{} 
 \section{Acknowledgements}

The author thanks the National Science Foundation for support through grant NSF-CHE-95-13558, CIRE 98-72689 and to the Department of Energy for grant DOE-EPSCoR grant DE-FCO2-91ER75674.

\end{document}